\documentclass[12pt,a4paper]{article}%

\pdfoutput=1

\usepackage{amsmath,amssymb,amsfonts,multicol}
\usepackage[normal,font=small,labelfont=bf,labelsep=period]{caption}
\usepackage[pdftex]{color,graphicx}
\usepackage[english]{babel}
\usepackage[compress]{cite}

\usepackage[dvipsnames]{xcolor}
\definecolor{darkgreen}{rgb}{0,0.65,0}
\usepackage{slashed}

\def\sb{\sin{\beta}}
\def\cb{\cos{\beta}}
\def\tb{\tan{\beta}}
\def\sa{\sin{\alpha}}
\def\ca{\cos{\alpha}}

\newcommand{\icol}[1]{% inline column vector
  \left(\begin{smallmatrix}#1\end{smallmatrix}\right)%
}

\begin{document}
\begin{flushright}
\footnotesize
{IFT-UAM/CSIC-16-008}\\
{FTUAM-16-3}
%SACLAY--T12/xxx
\end{flushright}
\color{black}
%\vspace{0.3cm}

\begin{center}
{\Large\color{Red}\bf Di-Photon excess in the 2HDM: hasting
  towards the instability and the non-perturbative regime\\[1mm] }

\medskip
\bigskip\color{black}\vspace{0.6cm}

{
{\large\bf Enrico Bertuzzo}\ $^a$,
{\large\bf Pedro A.N. Machado}\ $^{b,c}$,
{\large\bf Marco Taoso}\ $^{b,c}$
%{\large\bf Author 3}$^c$
}
\\[7mm]

{\it $^a$ Instituto de F\'isica, Universidade de S$\tilde{a}$o Paulo,\\ C.P. 66.318, 05315-970 S$\tilde{a}$o Paulo, Brazil.}\\[3mm]
{\it $^b$  Departamento de F\'isica Te\'orica, Universidad Aut\'onoma de Madrid,\\ Cantoblanco E-28049 Madrid, Spain.}\\[3mm]
{\it $^c$  Instituto de F\'isica Te\'orica UAM/CSIC,\\Calle Nicol\'as Cabrera 13-15, Cantoblanco E-28049 Madrid, Spain.}\\[3mm]
\end{center}

\bigskip

\centerline{\large\bf Abstract}
\begin{quote}
\color{black}\large We challenge the interpretation of the di-photon
excess recently observed by both ATLAS and CMS in a two Higgs doublet
framework. Due to the large enhancement necessary to obtain the
observed di-photon signal, a large number of colored and charged
vector-like fermions are called for.  We find that even before the hypercharge
gauge coupling becomes non perturbative, the one loop effects of these
fermions abruptly drive the scalar potential to instability.
\end{quote}

%\tableofcontents
\newpage

\section{Introduction}
\label{sec:introduction}

Recently, ATLAS~\cite{ATLAS} and CMS~\cite{CMS} have reported excesses
in the di-photon channel using Run 2 data at $\sqrt{s}=13$ TeV center
of mass energy.  The ATLAS collaboration, with an integrated
luminosity of 3.2 fb$^{-1,}$ has found an excess at a di-photon
invariant mass of $\simeq 750$ GeV, with a local significance of $3.9
\sigma$ (2.3$\sigma$ after the look-elsewhere effect). Using 2.6
fb$^{-1}$ of data, CMS on its turn has observed an excess peaking at
an invariant mass of 760~GeV, with a local significance of $2.6
\sigma$ ($1.2 \sigma$ global). Although there is a very mild
preference for a resonance width of about 45~GeV, the data is yet too
insipid to support any claim in this direction.

These intriguing results have been intensively investigated by the
community~\cite{papersexcess}. A simple possibility is that the signal
comes from the decay of a spin 0 or 2 (by virtue of the Landau-Yang
theorem) resonance decaying into a two photons final state. If the
resonance participate in the breaking of electroweak (EW) symmetry, a
two Higgs doublet model (2HDM) is one of the simplest scenario to
focus on~\cite{2HDMexcess}. In this scenario, the resonance can be
identified with the heaviest of the CP even scalars (H), with the
CP-odd state (A), or even with a superposition of both, in case their
masses are degenerate within $\mathcal{O}(10~{\rm GeV})$. 

Nevertheless, the fact that the decay of scalars to di-photons is loop
induced poses many difficulties in building up a compelling model to
explain the data. As nothing beyond ordinary was observed consistent
with a 750~GeV resonance decaying to other channels, like diboson or
$t\bar t$, the scalars couplings to standard model (SM) fermions and
massive gauge boson need to be quite suppressed, which in turn also
suppress the production via gluon fusion and the decay to
photons. Therefore, to explain the ATLAS and CMS excesses additional
field content is called for. The effective couplings of the resonance
to photons and gluons can be enhanced by introducing new vector-like
quarks or leptons. Still, to obtain enough enhancement it is necessary
to have a large number of VL fermions, electrically charged and
possibly colored, with sizeable couplings to the heavy scalars.  In
this work we investigate the consequences of such profusion of
particles for the stability of the EW vacuum, and the evolution of the
gauge couplings at high scales.

The paper is organized as follow: in Section~\ref{sec:2HDM} we
introduce the model and we show the region of the parameter space
where the excess can be accommodated. In Section~\ref{sec:running} we
perform an analysis of the Renormalization Group Equation (RGE)
evolution of the relevant couplings. Finally, we conclude in
Sec.~\ref{sec:conclusions}.

%% These scalars could be dominantly produced at the LHC in gluon fusion
%% processes.

%We show that the couplings of the extra fermions to the resonance
%induces a loss of perturbativity of the gauge couplings and
%destabilize the EW vacuum. These effects occurs at scale

%%%%%%%%%%%%%%%%%%%%%%%%%%%%
%%%%%%%%%%%%%%%%%%%%%%%%%%%%

\section{2HDM and vector-like fermions}
\label{sec:2HDM}

The most general 2HDM potential compatible with the gauge symmetries
of the Standard Model (SM) is the following (see
e.g.~\cite{Branco:2011iw}):

\begin{eqnarray}\label{eq:2HDMpotential}
V(\Phi_1,\Phi_2) &=&  m_{11}^2 \Phi_1^{\dagger} \Phi_1+m_{22}^2 \Phi_2^{\dagger} \Phi_2 -m_{12}^2( \Phi_1^{\dagger} \Phi_2+ \Phi_2^{\dagger} \Phi_1)+\frac{\lambda_1}{2} (\Phi_1^{\dagger} \Phi_1)^2 +\frac{\lambda_2}{2} (\Phi_2^{\dagger} \Phi_2)^2 \\
&+& \lambda_3 \Phi_1^{\dagger} \Phi_1  \Phi_2^{\dagger} \Phi_2 + \lambda_4 \Phi_1^{\dagger} \Phi_2  \Phi_2^{\dagger} \Phi_1 + \frac{\lambda_5}{2} [  (\Phi_1^{\dagger} \Phi_2)^2 + (\Phi_2^{\dagger} \Phi_1)^2]\, ,
\end{eqnarray}

\noindent with $\Phi_1$ and $\Phi_2$ two complex SU(2) doublets with
hypercharge 1/2. Here, we have assumed a $Z_2$ symmetry
($\Phi_{1}\rightarrow -\Phi_{1}$), except for the soft breaking term
proportional to $m_{12}^2.$ 
For each doublet we can define:

\begin{eqnarray}
\Phi_i &=&  \icol{\phi^+_i \\  (v_i+\rho_i+i\eta_i)/\sqrt{2}}, \hspace{1cm} i=1,2
\end{eqnarray}

\noindent  In the following we assume that the CP symmetry is respected by the scalar potential, so the vacuum-expectation values $v_1$ and $v_2$ are real numbers. Moreover $\sqrt{v_1^2+v_2^2}=v_{SM}=246.2$ GeV.
This scalar theory contains five massive states: a charged scalar $H^{\pm},$ a CP odd state $A$ and two CP even particles $h,H$, the lightest of those ($h$) is identified with the 125 GeV Higgs boson.
The mass eigenstates are obtained after performing the  following rotations: 

\begin{equation}
\begin{array}{lcl}
H =  \rho_1  \sa +  \rho_2 \ca\, , &\hspace{0.5cm}& A =   - \eta_1 \cb + \eta_2 \sb\, , \\
 h =  \rho_1 \ca  -  \rho_2 \sa\, ,  &\hspace{0.5cm}& H^+ = - \phi^+_1 \cb + \phi^+_2 \sb\, ,
\end{array}
\end{equation}

\noindent  where the angles $\beta$ and $\alpha$ are:

\begin{eqnarray}
\tb = \frac{v_1}{v_2}\, , \hspace{1cm} \tan{2 \alpha} = \frac{2 (-m_{12}^2 +\lambda_{345} v_1 v_2)}{m_{12}^2 (v_1/v_2-v_2/v_1)+\lambda_2 v_2^2 -\lambda_1 v_1^2}\, ,
\end{eqnarray}

\noindent  and $\lambda_{345}=\lambda_3+\lambda_4+\lambda_5.$

\noindent The quartic couplings of the scalar potential can be
expressed in terms of the masses of the scalars, $\tb$ and $M^2\equiv
\frac{m_{12}^2}{\sb \cb}:$

\begin{eqnarray}
\lambda_1 &=& \frac{1}{v^2} \left(  -\tan^2 \beta M^2 +\frac{\sin^2 \alpha}{\cos^2 \beta}m_h^2 +\frac{\cos^2 \alpha}{\cos^2\beta}m_H^2  \right)\, ,\\
\lambda_2 &=& \frac{1}{v^2} \left(  -\cot^2 \beta  M^2 +\frac{\cos^2\alpha}{\sin^2\beta}m_h^2 +\frac{\sin^2\alpha}{\sin^2\beta}m_H^2  \right)\, ,\\
\lambda_3 &=& \frac{1}{v^2} \left(  - M^2 +2m_{H^{\pm}}^2  +\frac{ \sin 2 \alpha}{\sin 2 \beta} (m_H^2-m_h^2)  \right)\, ,\\
\lambda_4 &=& \frac{1}{v^2} \left(  M^2 +m_A^2-2m_{H^{\pm}}^2  \right)\, ,\\
\lambda_5 &=& \frac{1}{v^2} \left(  M^2 -m_A^2  \right)\, .
\end{eqnarray}

The potential is unbounded from below if the following conditions are fulfilled:
\begin{eqnarray}
\lambda_{1,2}>0, \hspace{1cm} \lambda_3 > -(\lambda_1 \lambda_2)^{1/2} \hspace{1cm} \lambda_3+\lambda_4-|\lambda_5|> -(\lambda_1 \lambda_2)^{1/2}.
\label{eq:unbounded}
\end{eqnarray}
The unitarity constraints can be found in
ref.~\cite{Branco:2011iw}. As we will see later, the impact of the VL
fermions will be so large that the precise expressions for the
unitarity bound will not matter.

Concerning the couplings of the SM fermions to the physical scalars,
different configurations are possible. Here we consider the type I
2HDM, a scenario where all the SM fermions couple only to one doublet
($\Phi_1$ by convention).  Notice that the $Z_2$ symmetry of the
potential suits this case very well, as it can be the reason of why SM
fermions do not couple to $\Phi_2$. In this scenario, the couplings
between fermions, gauge bosons and the scalars, after rotating to the physical
basis, are given by
\begin{eqnarray}\label{eq:couplings}
  y^f_h &=& y^f_{SM} \left( \sin(\beta-\alpha) + \frac{\cos(\beta-\alpha)}{\tan\beta} \right),\\
  y^f_{H} &=& y^f_{SM}\left( \cos(\beta-\alpha) - \frac{\sin(\beta-\alpha)}{\tan\beta} \right),\\
  y^f_A &=& \pm \, y^f_{SM}\frac{1}{\tan\beta},\\
  g_{hVV} &=& 2 \sin(\beta-\alpha) \frac{m_V^2}{v}, \\
  g_{HVV} &=& 2 \cos(\beta-\alpha) \frac{m_V^2}{v},
\end{eqnarray}
where $y^f_{h,A,H,SM}$ are the Yukawa coupling of the fermion $f$ to the
scalars $h$, $A$, $H$, and the standard model Yukawa,
respectively. The $\pm$ sign in $y^f_A$ applies to up and down quarks, respectively. In order to have SM-like couplings for the lightest scalar, we must invoke the so called ``alignment limit'', $\beta-\alpha \sim \pi/2$. As can be seen, in this limit the couplings of all the heavy scalars are suppressed.

As explained in the Introduction, additional matter is necessary to
reproduce the di-photon excess. Following the recent literature~\cite{2HDMexcess}, we
introduce new vector-like (VL) quarks. The minimal scenario includes
one VL SU(2) doublet (Q) and one SU(2) singlet (D) with an appropriate
hypercharge assigments. We take the VL in the fundamental
representation of SU(3).  The lagrangian includes:

\begin{equation}\label{eq:lagrangian}
\mathcal{L} \supset  y_i ^Q\bar{Q}_R \Phi_i D_L + y_i^D \bar{Q}_L \Phi_i D_R. 
\end{equation}

For simplicity we take $y_i^Q=y_i^D \equiv y_i$ and a common VL mass
$M_{VLQ}$ for these extra fermions.

\subsection{Signal and constraints}

To account for the ATLAS and CMS excess, the total cross-section  in di-photons at 750 GeV should be~\cite{ATLAS,CMS}

\begin{eqnarray}\label{eq:signal}
\sigma(pp\rightarrow \gamma\gamma)_{ATLAS}=(10\pm 3)\, \mbox{fb} \hspace{1cm} \sigma(pp\rightarrow \gamma\gamma)_{CMS}=(6\pm 3) \, \mbox{fb}\,.
\end{eqnarray}

We have computed the production cross-section of $H$ and $A$ in gluon fusion processes and we have included decays of those resonances into SM fermions and gauge bosons.
We get:

\begin{eqnarray}
\sigma(H/A)_{\gamma\gamma}=\frac{C_{gg}}{M s} \Gamma_{gg}\mbox{ } BR(H/A\rightarrow \gamma\gamma)
\end{eqnarray}

\noindent The mass of the resonance $M$ is fixed at 750 GeV and $\Gamma_{gg}$ is the width in gluons. The partonic  integral factor $C_{gg}$ is obtained employing the set of pdf MSTW2008NLO~\cite{Martin:2009iq} at a scale $\mu=M,$ and it reads $C_{gg}=2137$ at  $\sqrt{s}=13$ TeV.
In the following we assume that $H$ and $A$ are degenerate in mass, in such a way that they both contribute to the di-photon signal.

A viable scenario to explain the excess can be obtained with the following configuration:

\begin{itemize}
\item[$\diamond$] the 2HDM should be close to the alignment limit, $|\cos (\beta-\alpha)| \lesssim 0.3.$ This is necessary to reproduce the observed couplings of the mostly SM Higgs $h$ to gauge bosons, Eq.~(\ref{eq:couplings}). The decays of $H$ to gauge bosons are suppressed, since they are proportional to $\sin (\beta-\alpha);$

\item[$\diamond$] In the 2HDM type I, the couplings of $H$ and $A$ to the SM fermions are universally suppressed by $1/\tb,$ compared to the SM  couplings to the Higgs boson. Therefore, focusing on large $\tb,$ one can suppress the decays of $H$ and $A$ into SM fermions. Moreover, this condition is necessary to minimize the departure of the $h$ couplings to the SM values;

\item[$\diamond$] In the case of large $\tan\beta$, the contribution
  to $h \to \gamma \gamma$ due to the VL quarks of
  Eq.~(\ref{eq:lagrangian}) are proportional to $y_1^2$, while the
  contributions to $(H,A) \to \gamma\gamma$ are proportional to the
  product $y_1 y_2$~\cite{2HDMexcess}. As such, in order to keep under
  control the deviations of the $h$ couplings to gluons and photons
  induced by the VL quarks, a hierarchy $y_1 <y_2$ should be imposed.
  \end{itemize}

\begin{figure}[!t]
\begin{center}
\includegraphics[width= 0.48 \textwidth]{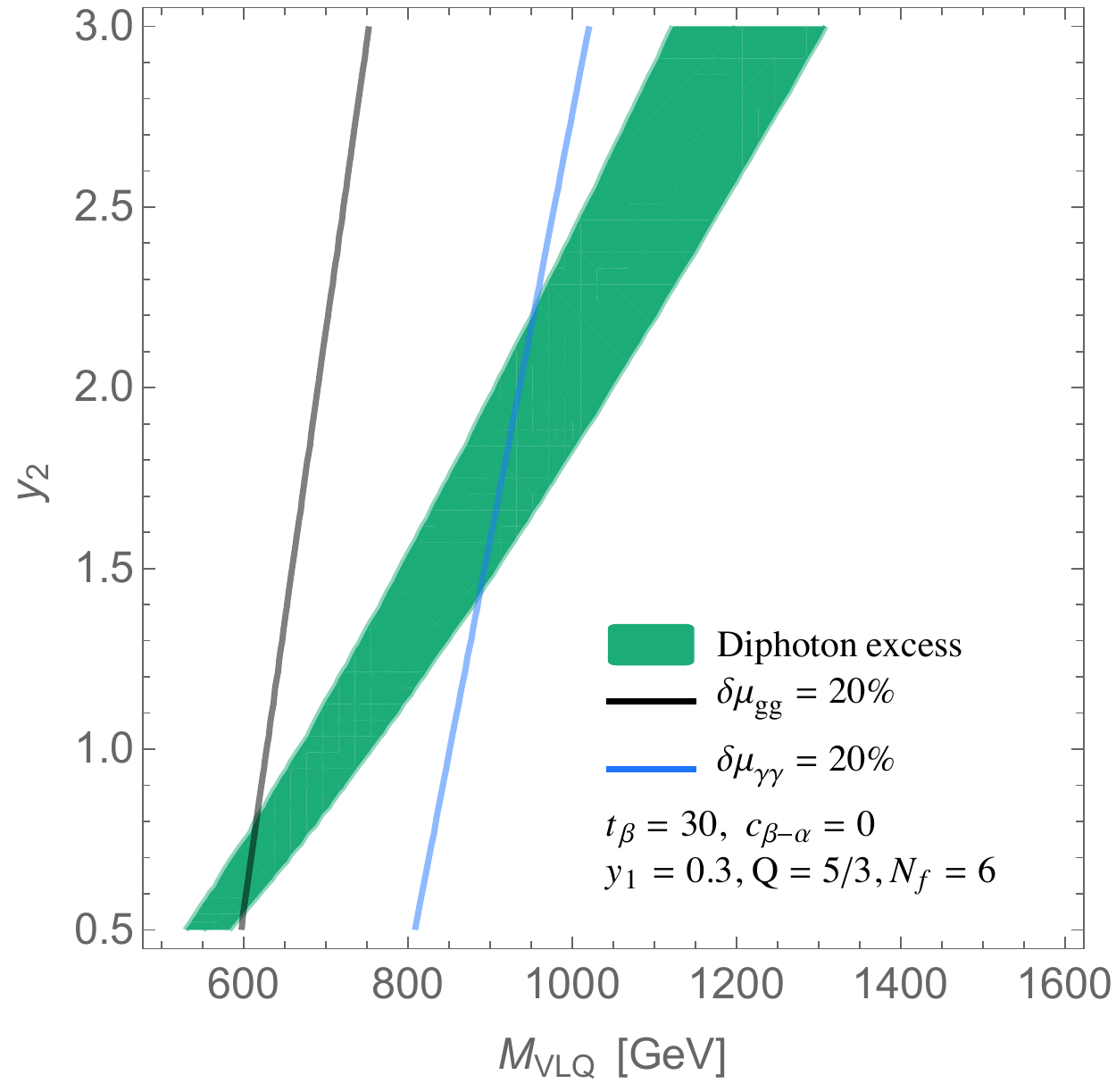} \quad
\includegraphics[width= 0.48 \textwidth]{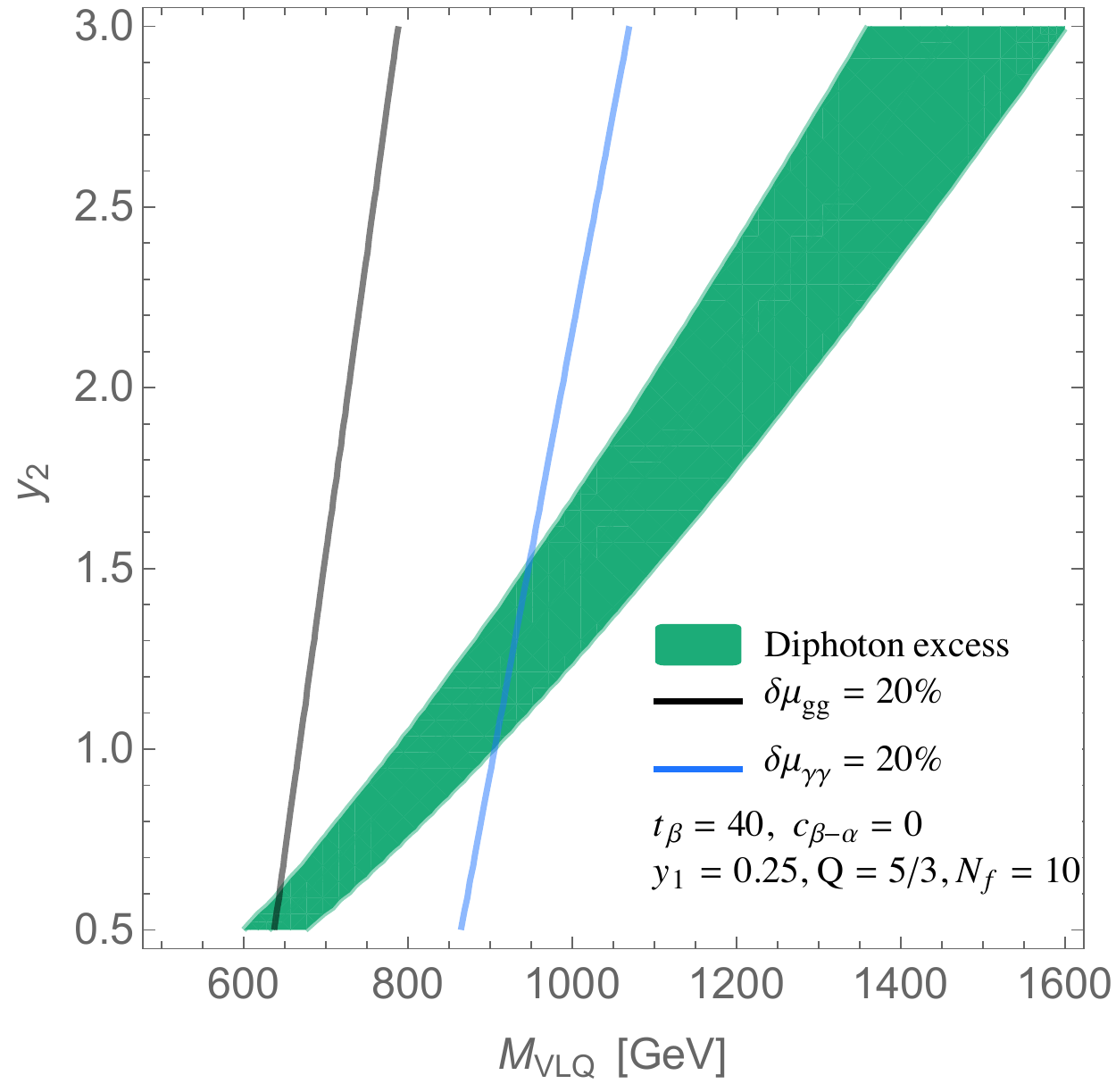}
\caption{\em \small \label{fig:xsect} {\bfseries } Left plot:
  $\tb=30,$ $y_1=0.3,$ $Q=5/3$ and $N_f=6$. Right plot: $\tb=40,$
  $y_1=0.25,$ $Q=5/3$ and $N_f=10$. The region on the left of the black
  and blue lines is excluded respectively by measurements of the $h$
  couplings to gluons and photons. Inside the green region the
  $\sigma(pp\rightarrow (H/A)\rightarrow \gamma\gamma)$ can explain
  the di-photon excess.}
\end{center}
\end{figure}

In Fig.~\ref{fig:xsect} we show some benchmark cases. In the green
area, $\sigma(H/A)_{\gamma\gamma}=(3\div10)$ fb and the di-photon
excess can be reproduced. The regions on the left of the blue and
black lines are excluded since the deviations of $h$ couplings to
$\gamma\gamma$ and $gg$ are larger than
$20\%$~\cite{hcouplings,CMS:2015kwa}.  As can be noticed from the left
panel of Fig.~\ref{fig:xsect}, the measurements of the $h$ couplings
strongly constrain this model, and a consistent explanation of the
di-photon excess requires a quite extreme configuration. Moreover, the
direct searches for VLQ through the decay processes $T^{+5/3} \to W^+
t$, $T^{+2/3}\to bW^+$, $B^{-1/3}\to t W^-$ and $B^{-1/3}\to b h$
constrain the mass of these fermions to be above
800~GeV~\cite{Chatrchyan:2013wfa}, 705~GeV~\cite{CMS:1900uua},
800~GeV~\cite{CMS:2014cca}, and 846~GeV~\cite{CMS:2014afa},
respectively.~\footnote{These bounds were derived assuming only one VL
  quark at a time. A large multiplicity $N_f$ of VL quarks with
  similar masses would na\"ively raise the cross sections under
  consideration by $N_f$, and therefore the bounds would typically be
  at the TeV scale.}

For instance, we have explicitly checked that, in order to explain Eq.~(\ref{eq:signal}) without incurring into troubles with the direct bounds, for VL quarks with charge $Q=5/3$ more than $N_f \gtrsim 5$ families are required. Such large multiplicity suggests
that the new states could dramatically modify the evolutions of the
couplings of the theory, through RGE effects.  This is the focus of
the next section.

\section{RGE running and the fate of the EW vacuum}
\label{sec:running}

Let us now study the RGE evolution for the SM gauge couplings ($g_3,$
$g^{\prime}$, $g$), the top Yukawa coupling ($y_t$) and the quartic couplings
$\lambda_i$ of the 2HDM potential, Eq.~(\ref{eq:2HDMpotential}). The corresponding 1-loop $\beta$ functions are
reported in Appendix~\ref{appendix}.  We have approximated the Coleman-Weinberg effective potential with the RGE-improved tree level potential, 
substituting the bare couplings with the corresponding RGE running quantities.
%In order to compute the contribution of the VL fermions to the $\beta$ functions of the quartic couplings, we have computed the Coleman-Weinberg effective potential, substituting the bare couplings with the corresponding RGE running quantity. 
In this way, the stability of the EW vacuum is still given by Eqs.~(\ref{eq:unbounded}).

\begin{figure}[!t]
\begin{center}
\includegraphics[width= 0.48 \textwidth]{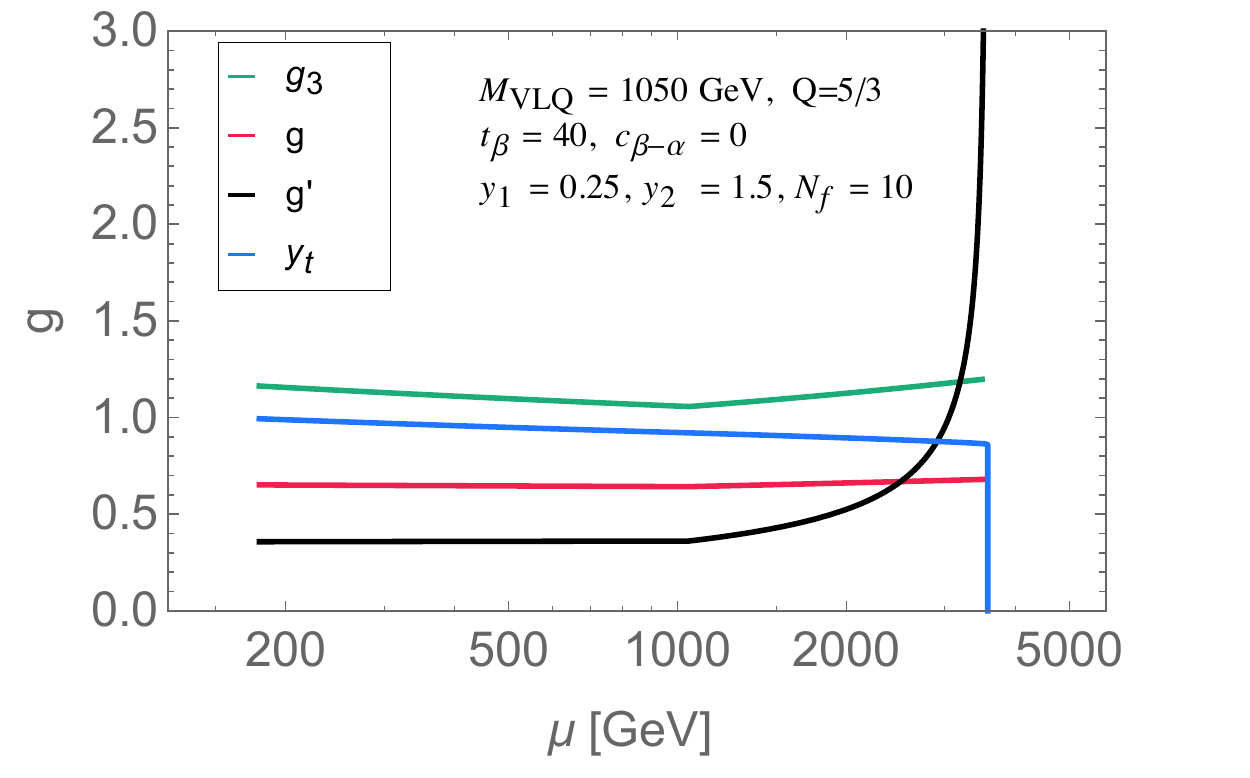} \quad
\includegraphics[width= 0.48 \textwidth]{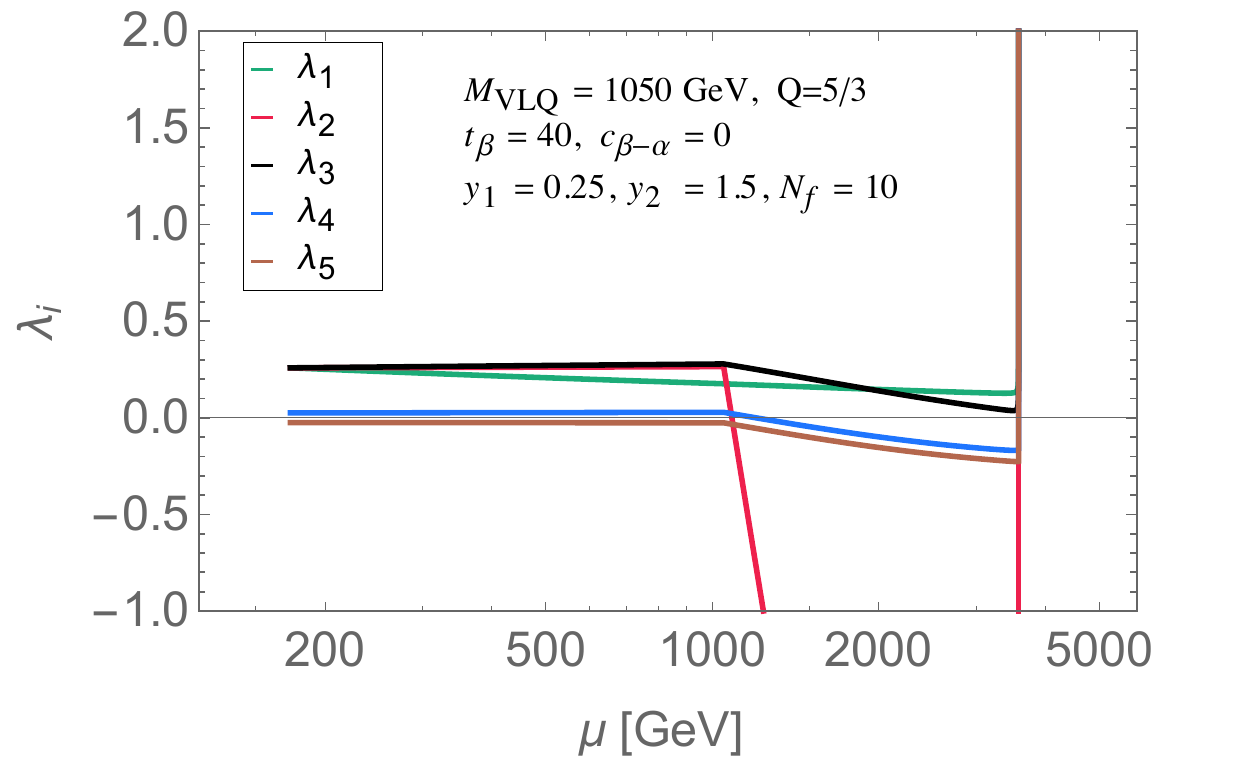}
\caption{\em \small \label{fig:RGE} {\bfseries } Left panel: RGE evolution of $g_3,$ $g^{\prime},$ $g$ and $y_t$ respectively in green, red, black and blue lines. Right panel: RGE evolutions of $\lambda_i.$ For these plots, we have chosen $M_{VLQ}=1050$ GeV, $y_1=0.25,$ $y_2=1.5,$ $Q=5/3$ and $N_f=10.$ The masses of the scalar sector are $M_H=750$ GeV, $M_A=751$ GeV and $M_{H^{\pm}}=750$ GeV.}
\end{center}
\end{figure}

Our results are shown in Fig.~\ref{fig:RGE}. To fix the boundary conditions for the quartic couplings we must choose a possible scalar spectrum. For simplicity, and in order to minimize the contribution to precision observables~\cite{Branco:2011iw}, we take all the heavy scalars to be degenerate at $M_H\simeq M_A \simeq M_{H^{\pm}}=750$ GeV, and we fix $m_h = 125$ GeV. We expect our results to be valid in general, and not only for this choice of boundary conditions. Having fixed the boundary conditions, we can now evolve the couplings to high energy, fixing the parameters in the VL quark sector in such a way to explain the di-photon excess, Fig.~\ref{fig:xsect}. As an example, we fix a common VL quark bare mass at $M_{VLQ} = 1050$ GeV, with charge $Q=5/3$ and couplings $y_1 = 0.25$ and $y_2 = 1.5$. We also fix the number of VL families to $N_f = 10$. As can be seen from Eq.~(\ref{eq:beta_func}), the hypercharge gauge coupling receives a large positive contribution which rapidly drives it to non perturbative values slightly above the VL threshold, as confirmed in Fig.~\ref{fig:RGE}.~\footnote{Similar results have been obtained in the case of a scalar singlet~\cite{Son:2015vfl}.}  However, another important effect can be inferred from the RGE's in Eq.~(\ref{eq:beta_func}). Indeed, $\lambda_2$ receives a large negative contribution proportional to $N_f \, |y_2|^4$, which drives it to negative values slightly above the VL quark threshold. Comparing with the bounds in Eq.~(\ref{eq:unbounded}), we see that this makes the potential unbounded from below already at the TeV scale, before the theory reaches a non perturbative regime.

These findings imply that, in this scenario, new physics should occur
at the scale of the VL quarks to stabilize the scalar
potential. Moreover the model becomes strongly coupled around these
energies, at which new degrees of freedom should emerge and a new
description of this theory is necessary.  Although, we have not
performed a complete scan of the parameter space of the model, we
expect our conclusions to be generic for simple 2HDM interpretations
of the diphoton excess. Indeed, i) generically large multiplicity and
couplings of the VL quarks are needed to explain the excess, ii) this
implies large corrections of the RGE evolutions of the gauge
couplings, independently on the choice of $\lambda_i,$ and iii)
starting from a weakly-coupled theory some of the $\lambda_i $
unavoidably runs toward negative values, before being attracted to
large positive values by the contribution of gauge couplings.

\section{Conclusions}
\label{sec:conclusions}

Although still not statistically relevant, it is tantalizing to
interpret the recently observed di-photon excess at $750$ GeV in terms
of extensions of the Standard Model. In this paper we have focused on
the case of the 2HDM. As is well known~\cite{2HDMexcess}, a 2HDM alone
cannot reproduce the observed signal, calling thus for a more
elaborate SM extension. The simplest possibility, already considered
in the literature, is the one of a 2HDM augmented with vector-like
quarks. In principle, such a framework could explain the excess
through the decays of the heavy $H$ and/or $A$ scalars. Here, we have
shown that a consistent realization of this scenario present severe
difficulties. Our results can be summarized as follows: a large number
of VL quarks is required in order to get the correct cross-section
into two photons.  The presence of such extra matter strongly affects
the RG evolution of the hypercharge gauge coupling, which rapidly
reaches non perturbative values a few hundreds of GeV above the VL
quarks threshold. However, even before this happens, some of the Higgs
quartic couplings become negative, destabilizing the scalar potential,
as it becomes unbounded from below. We thus reach the following broad
conclusions: new physics is required around the VL quarks threshold in
order to stabilize the vacuum, and a strongly coupled description of
the theory should emerge at slightly larger scales. Let us now comment
on what can be inferred about possible UV completions of this
scenario. A natural possibility would be to consider a composite 2HDM
scenario~\cite{Mrazek:2011iu,Bertuzzo:2012ya,Sanz:2015sua,Antipin:2015jia}~\footnote{In
  principle, a composite 2HDM could also emerge from a fermionic UV
  completion of the Minimal Composite Higgs
  Model~\cite{vonGersdorff:2015fta}.}, in which vector like quarks are
naturally expected around the TeV scale. However, in order to
stabilize the vacuum, additional bosonic degrees of freedom are needed
around the same scale. Whether such additional particles should be
part of an extended coset, or can be additional vectorial resonances,
is still an open question that we may explore in a forthcoming work.

\newpage

\small
\subsubsection*{Acknowledgments}

\footnotesize
\noindent Funding and research infrastructure acknowledgements: 
\begin{itemize}
\item[$\ast$] Centro de Excelencia Severo Ochoa Programme SEV-2012-0249,
\item[$\ast$] MINECO through a Severo Ochoa fellowship with the Program SEV-2012-0249,
\item[$\ast$] FPA2015-65929-P and Consolider MultiDark CSD2009-00064,
\item[$\ast$] Conselho Nacional de Ci\^encia e Tecnologia (CNPq),
\item[$\ast$] ITN INVISIBLES (PITN-GA-2011-289442)
\end{itemize}

\bigskip
\appendix

\section{1-loop $\beta$ functions}
\label{appendix}

We now list the relevant 1-loop  beta functions of the model:

\begin{equation}\label{eq:beta_func}
\begin{array}{rcl}
\beta_{g^{\prime}} &=&\beta_{g^{\prime}}^{2HDM}+\frac{4 N_f}{3} \frac {g^{\prime 3}}{16 \pi^2}  (9 Q^2 + 6 Q +3/2) ,\\
\beta_{g} &=&\beta_{g}^{2HDM}+2 N_f \frac {g^{ 3}}{16 \pi^2}, \\
\beta_{g_3} &=&\beta_{g_3}^{2HDM}+2 N_f \frac {g_3^{ 3}}{16 \pi^2}, \\
\beta_{y_t} &=&\beta_{y_t}^{2HDM}, \\
\beta_{\lambda_1} &=& \beta_{\lambda_1}^{2HDM} -3 N_f  \frac{  |y^{Q}_{1}|^{4}+|y_{1}^{D}|^{4}}{4 \pi^2}, \\
\beta_{\lambda_2} &=& \beta_{\lambda_1}^{2HDM} -3 N_f  \frac{   |y^{Q}_{2}|^{4}+|y_{2}^{D}|^{4}}{4 \pi^2}, \\
\beta_{\lambda_3} &=& \beta_{\lambda_3}^{2HDM} -3 N_f  \frac{   |y^{Q}_{1}|^{2}  |y^{Q}_{2}|^{2}+|y_{1}^{D}|^{2} |y_{2}^{D}|^{2}   }{4 \pi^2}, \\
\beta_{\lambda_4} &=& \beta_{\lambda_4}^{2HDM} -3 N_f  \frac{   |y^{Q}_{1}|^{2}  |y^{Q}_{2}|^{2}+|y_{1}^{D}|^{2} |y_{2}^{D}|^{2}   }{4 \pi^2}, \\
\beta_{\lambda_5} &=& \beta_{\lambda_5}^{2HDM} -3 N_f  \frac{   ( y^{Q *}_{1}   y^{Q}_{2} )^{2}+(y_{1}^{D} y_{2}^{D *} )^{2}   }{4 \pi^2}, \\
\beta_{\lambda_6} &=& \beta_{\lambda_6}^{2HDM} -3 N_f  \frac{    y^{Q *}_{1}   y^{Q}_{2}  |y^{Q}_{1}|^{2} +y_{1}^{D} y_{2}^{D *}  |y^{D}_{1}|^{2}   }{4 \pi^2}, \\
\beta_{\lambda_7} &=& \beta_{\lambda_7}^{2HDM} -3 N_f  \frac{    y^{Q *}_{1}   y^{Q}_{2}  |y^{Q}_{2}|^{2} +y_{1}^{D} y_{2}^{D *}  |y^{D}_{2}|^{2}   }{4 \pi^2}.\\
\end{array}
\end{equation}

\smallskip

\noindent The $\beta$ functions of the 2HDM, $\beta^{2HDM},$ can be found in~\cite{Branco:2011iw}. We have included in the $\beta$ functions the dominant contribution of the VL quarks from the renormalization of the 4-point scalar vertexes, while we have disregarded the sub-leading contribution from the wave-function renormalization.
Also note that $Z_2$ breaking terms proportional to $\lambda_6$ and $\lambda_7$ (see~\cite{Branco:2011iw} for their definition) are radiatively generated.

\newpage

\footnotesize
\begin{multicols}{2}
  
\end{multicols}

\end{document}